\documentclass{elsarticle}

\usepackage{adjustbox}
\usepackage[linesnumbered,ruled,vlined]{algorithm2e}
\usepackage{algorithmic}
\usepackage{amsfonts}
\usepackage{amsmath}
\usepackage{amssymb}
\usepackage{amsthm}
\usepackage{array}
\usepackage{bm}
\usepackage{booktabs}
\usepackage{caption}
\usepackage{colortbl}
\usepackage{enumitem}
\usepackage{float}
\usepackage[T1]{fontenc}
\usepackage{graphicx}
\usepackage{hyperref}
\usepackage{latexsym}
\usepackage{listings}
\usepackage{inconsolata}
\usepackage[utf8]{inputenc}
\usepackage{makecell}
\usepackage{marvosym}
\usepackage{microtype}
\usepackage{multirow}
\usepackage{subcaption}
\usepackage{tabularx}
\usepackage{tikz}
\usetikzlibrary{matrix,shapes,arrows,fit}
\usepackage{times}
\usepackage{xcolor}

\journal{Science of Computer Programming}

\definecolor{deepred}{rgb}{0.8, 0, 0} 
\definecolor{deepgreen}{rgb}{0, 0.6, 0} 
\definecolor{deepblue}{rgb}{0, 0, 0.8} 

\definecolor{code_background}{rgb}{252,252,252}
\definecolor{custorange}{rgb}{0.898, 0.376, 0.02}

\lstdefinestyle{cstyle}{
    language=C,
    basicstyle=\fontsize{8pt}{8pt}\ttfamily,
    keywordstyle=\color{blue},
    stringstyle=\color{orange},
    commentstyle=\color{custorange},
    backgroundcolor=\color{code_background},
    numbers=left,
    numberstyle=\fontsize{8pt}{8pt},
    numbersep=2pt,
    frame=tb,
    rulecolor=\color{black},
    framerule=0.5mm,
    framesep=2mm,
    xleftmargin=1.0mm,
}

\begin{document}

\begin{frontmatter}

\title{Enhancing Automated Loop Invariant Generation for Complex Programs with Large Language Models}

\author[sjtu]{Ruibang Liu}
\author[sjtu]{Minyu Chen}
\author[sjtu]{Ling-I Wu}
\author[sjtu]{Jingyu Ke}
\author[sjtu]{Guoqiang Li\corref{cor}}\ead{li.g@sjtu.edu.cn}
\cortext[cor]{Corresponding author.}
\affiliation[sjtu]{organization={Shanghai Jiao Tong University},
            postcode={200240}, 
            state={Shanghai},
            country={China}}

\begin{abstract}
Automated program verification has always been an important component of building trustworthy software. While the analysis of loops remains a theoretical challenge, the automation of loop invariant analysis has effectively resolved the problem. However, existing invariant generation tools are predominantly effective for programs with purely numerical or purely pointer-based structures. Real-world programs often mix complex data structures and control flows. These structures can include arrays, pointers, and recursive definitions, while control flows may involve multiple nested or concurrent loops. Traditional methods generally only generate invariants for simple numerical programs or specific segments, lacking broad applicability. In order to automatically generate loop invariants for real-world programs, we proposed \textit{ACInv}, an Automated Complex program loop Invariant generation tool, which combines static analysis with prompting with Large Language Models (LLM) to generate the proper loop invariants. We employ static analysis to systematically decompose the program's data structures and loops. This involves layer-by-layer transmission of structural information about variables, numerical data, and the complete loop structure to the LLM, enabling the generation of corresponding invariants. In comparison to prior work on AutoSpec, we delve deeper into the variable information within each loop. We conducted experiments on ACInv, which showed that ACInv outperformed previous tools on data sets with data structures and maintained similar performance to the state-of-the-art tool AutoSpec on numerical programs without data structures. For the total data set, ACInv can solve 21\% more examples than AutoSpec, and can generate reference data structure templates.
\end{abstract}


\end{frontmatter}
\section{Introduction}

Program verification has progressively emerged as a fundamental component in the development of reliable software within the industry, thereby necessitating advanced tools for automated program verification. Nonetheless, the verification of program properties in the presence of loops is inherently undecidable. Within the framework of Hoare logic~\cite{10.1145/363235.363259}, \textit{loop invariants}, which are usually assertions that are always true during the entry and execution of the loop, serve as critical abstractions of loop properties. These invariants are indispensable in enabling verification tools to rigorously establish the correctness of programs. However, in the practical implementation of program verification tools such as CPAchecker~\cite{10.1007/978-3-642-22110-1_16}, CBMC~\cite{10.1007/978-3-642-54862-8_26}, SMACK~\cite{10.1007/978-3-319-08867-9_7} and Frama-C~\cite{10.1007/s00165-014-0326-7}, the derivation of loop invariants often necessitates manual intervention by domain experts, which poses a substantial impediment to the full automation of program verification.

Loop invariants are usually defined inductively, which means that they must meet the following two conditions: (1) \textbf{Initialization}: It is true before the first iteration of the loop; (2) \textbf{Maintenance}: If the invariant holds true at the beginning of an iteration, it must also hold true at the end of that iteration. From a theoretical perspective, for any real-world program containing loops, irrespective of the programming language or the complexity of the program's code structure or data structure, we aim to automatically generate correct loop invariants. Moreover, these loop invariants should be strong enough to support the verification of relevant properties under various conditions.

Although there is a substantial body of prior work on the automatic generation of loop invariants, these methods often face challenges when applied to the verification of real-world programs, primarily due to their generality and associated performance overhead. Abstract interpretation baesd methods~\cite{10.1007/978-3-540-27864-1_21, RODRIGUEZCARBONELL200754, 10.1007/978-3-319-46520-3_30} was initially adopted by researchers. These approaches typically employ static analysis techniques, such as value analysis and interval analysis, to trace the values manipulated by the program and subsequently synthesize these values to construct loop invariants. However, these methods are limited to analyzing programs with simple numerical types and are less effective in handling complex loops, advanced data types and structures, and fail to adequately address intricate relationships between variables. Template-based methods~\cite{10.1007/978-3-540-27864-1_7, 10298567, 10.1007/978-3-031-64626-3_19, 10.1007/978-3-540-45069-6_39, 10.1145/3563295} typically involve predefining invariants using SMT formulas, proposing parameterized invariant templates via uninterpreted functions or inequalities, and employing heuristic strategies to determine the parameters. While this approach can reveal more complex relationships between variables, it remains largely confined to linear numerical inequalities and imposes significant computational costs due to the introduction of SMT constraint solving. The widespread adoption of machine learning has led many researchers to propose learning-based methods~\cite{Ryan2020CLN2INV:, 10.1007/978-3-319-08867-9_5, NEURIPS2018_65b1e92c, 10.1145/3597926.3598047}, which infer suitable invariant templates through specific deductions and learn the corresponding parameters using various learning techniques to generate invariants. However, their effectiveness remains limited by the predefined templates and the quality of the training data. 

In recent years, \textit{Large Language Models} (LLMs), such as ChatGPT, have emerged as powerful tools for automating various generative tasks. Several studies~\cite{10.1145/3649828, 10.1007/978-3-031-64626-3_22} have demonstrated their strong performance in program-related reasoning tasks. Notably, recent works~\cite{chakraborty-etal-2023-ranking, pmlr-v202-pei23a, 10.1007/978-3-031-65630-9_16, liu2024generalloopinvariantgeneration, kamath2023findinginductiveloopinvariants} have highlighted the potential of LLMs in generating loop invariants. Trained on vast datasets encompassing both code and natural language, LLMs possess a sophisticated understanding of code patterns and semantic relationships. They excel at recognizing complex structures, such as loops, and are capable of automating invariant generation across extensive codebases, significantly reducing the need for manual effort.

Despite the potential of LLMs in invariant generation, several challenges remain. First, as highlighted by prior studies~\cite{10.1145/3560815, NEURIPS2022_9d560961, NEURIPS2023_271db992, yang2024large}, the inferential capabilities of LLMs require further exploration, particularly through carefully crafted prompts. Second, LLMs may produce incorrect invariants, and these errors can compound when the models are repeatedly invoked. To mitigate this issue, researchers have integrated SMT solvers or verification tools into the workflow to filter out erroneous invariants. However, this solution introduces significant performance overhead, and its effectiveness hinges on the strength of the verification tool. Moreover, certain correct yet complex loop invariants may be rejected by the tool, necessitating additional manual intervention, which conflicts with the goal of an end-to-end approach. Lastly, even when error-free, the invariants generated by LLMs may lack sufficient strength to capture precise properties, thereby failing to assist in verifying the target property.

To address these challenges, we propose \textit{ACInv}, an Automated Complex program loop Invariant generation tool leveraging Large Language Models (LLMs) in conjunction with static analysis. In order for ACInv to be able to abstract data structures and reason about invariants, we need some necessary program information. The tool begins by extracting detailed information about the program’s data structures and variable values within loops using static analysis techniques. ACInv then queries LLMs to generate appropriate predicate templates, which abstract potential data structure properties in the program. After this step, it inserts the predicates generated by LLMs as comments in the program, further analyzing variable information within loops and annotating those variables that undergo value changes in each iteration. For nested loops, the tool recursively queries LLMs, starting with the innermost loop. 

In order to enable ACInv to determine the correctness of the invariants it generates and further enhance the invariants it generates, we check and enhance each invariant generated by ACInv. This step is completed by the evaluator and optimizer built by LLM. The evaluator first evaluates the correctness of the invariant and then passes the evaluation results and the corresponding invariant to the optimizer for processing. For valid invariants, the optimizer seeks to refine them, while incorrect invariants are either weakened to align with the correct program states or discarded altogether.

We contend that \textit{ACInv} offers several distinct advantages over existing tools. First, it is capable of analyzing a wide range of complex, real-world code, with its sophisticated template processing for data structures and effective handling of nested loops, enabling it to tackle intricately structured programs. Second, \textit{ACInv} delivers a fully automated, end-to-end solution for loop invariant generation, significantly reducing the need for manual intervention throughout the process. Lastly, while maintaining a competitive level of correctness, \textit{ACInv} minimizes time overhead and optimizes the quality of the generated invariants.

To evaluate ACInv, we conducted a series of experiments. We conducted experiments on ACInv on datasets with and without data structures. The experiments showed that ACInv performed better than all current tools on datasets with data structures. On datasets without data structures, ACInv also performed close to the performance of the state-of-art tool AutoSpec~\cite{10.1007/978-3-031-65630-9_16}. Overall, ACInv can solve 21.08\% more examples than AutoSpec.

In summary, our contribution is:
\begin{itemize}
    \item We have built a better automatic invariant generation analysis method based on LLM that can handle relatively complex data structures while maintaining a relatively high correctness rate with low local time consumption, and built a tool ACInv based on it.
    \item We generated some predicate templates that may be useful for some commonly used data structures and other user-defined data structures on benchmarks such as ~\cite{liu2024generalloopinvariantgeneration, LIG-MM-repo} and SV-COMP~\cite{10.1007/978-3-031-57256-2_15, SV-comp-24-repo}.
\end{itemize}

\section{Background}
\subsection{Loop Invariant}
To provide a formal characterization of the invariant, we first abstract the loop and its invariants using Hoare logic~\cite{10.1145/363235.363259} as follows:
\begin{align*}
    \frac{\{I \wedge B\}\ S\ \{I\}}{\{I\}\ \mathbf{while}\ (B)\ \mathbf{do}\ (S)\ \{I \wedge \neg B\}}
\end{align*}

Where we abstract the general loop structure in the form of pseudo-code into the expression $\mathbf{while}\ (B)\ \mathbf{do}\ (S)$, defining $B$ as the guard condition, $S$ as the loop body, and $I$ as the sought loop invariant. Given a loop invariant $I$, if the body $S$ preserves $I$ under the condition that both $I$ and $B$ hold, i.e., $\{I \wedge B\}\ S\ \{I\}$ is valid, then initiating the loop in a state satisfying $I$ ensures that upon termination, the state satisfies both the invariant $I$ and the negation of the guard condition $\neg B$, formally, $\{I\}\ \mathbf{while}\ (B)\ \mathbf{do}\ (S)\ \{I \wedge \neg B\}$. This rule underscores the critical role of identifying a suitable loop invariant $I$ for verifying loop behavior.

\begin{figure}[!htbp]
    \centering
    \includegraphics[width=0.5\textwidth]{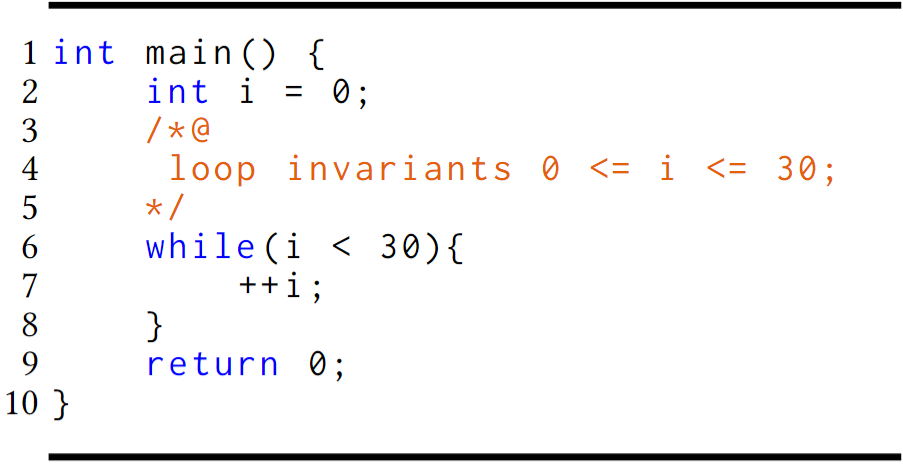}
    \caption{A simple C code example with loop invariants.}\label{fig1}
\end{figure}

\textbf{An example of loop invariants in ACSL} The vast majority of invariant research is based on the C language, so this paper will also use C language code as the research object. In the following paper, we will use ANSI/ISO C Specification Language (ACSL) to describe the loop invariants of the C program, which can be automatically checked by the verification tool Frama-C~\cite{10.1007/s00165-014-0326-7} and its plugin \textit{WP}. In ACSL, all the information needed is implemented as the annotation which begins with \verb|\\@| or is wrapped with \verb|\*@...*\|, where loop invariants are annotated beginning with \verb|loop invariant|.

Consider the example depicted in Fig~\ref{fig1}. The provided C code snippet above demonstrates a simple while loop. The program starts by declaring an integer variable \verb|i| and initializing it to \verb|0|. The \verb|while| loop runs as long as the condition \verb|i < 30| holds true. During each iteration, the value of \verb|i| is incremented by \verb|1| (\verb|++i|). When \verb|i| reaches \verb|30|, the condition \verb|i < 30| becomes false, and the loop terminates. The C program implements a bounded iteration construct with a formally verified loop invariant $\mathtt{0 \leq i \leq 30}$ specified in ACSL annotations, where the loop initializes $\mathtt{i}$ to $0$ and increments it via pre-increment operation $\mathtt{++i}$ until the guard condition $\mathtt{i < 30}$ becomes false. Through Hoare logic verification, the invariant satisfies initialization at entry ($\mathtt{i=0} \Rightarrow \mathtt{0 \leq 0 \leq 30}$), preservation under execution ($\mathtt{0 \leq i \leq 30} \land \mathtt{i < 30} \vdash \mathtt{0 \leq i+1 \leq 30}$ since $\mathtt{i \leq 29}$), and conclusion upon termination ($\mathtt{0 \leq i \leq 30} \land \neg \mathtt{(i < 30)} \equiv \mathtt{i = 30}$).

What we still need to add and explain is, the invariant and predicate referenced herein, along with those mentioned subsequently, are incorporated into the original code in the form of annotations. However, these annotations differ significantly from conventional comments. Traditional comments are authored by programmers to articulate their design intentions informally. In contrast, predicates and invariants are employed to specify certain program properties rigorously; they serve to assist in program verification and similar tasks, and are expressed in a strict formal language based on the ACSL (ANSI/ISO C Specification Language). Consequently, they serve distinct purposes and are conventionally demarcated by specialized syntactic markers in practice.

\subsection{Large Language Models (LLMs)}
In recent years, the field of Natural Language Processing (NLP) has achieved significant advances. State-of-the-art models, particularly those built on the transformer architecture~\cite{NIPS2017_3f5ee243}, have achieved remarkable success in addressing a wide range of natural language tasks. Large Language Models (LLMs), sophisticated neural networks trained on massive text corpora, have emerged as powerful tools for understanding and generating human-like language. By utilizing deep learning frameworks such as transformers, LLMs can effectively capture and model intricate linguistic patterns and relationships. These models represent a major leap forward in NLP, driven by the convergence of novel architectures, vast datasets, and enhanced computational resources. The development of large-scale pre-trained models like BERT (Bidirectional Encoder Representations from Transformers)~\cite{devlin-etal-2019-bert} and GPT (Generative Pre-trained Transformer)~\cite{NEURIPS2020_1457c0d6} has led to substantial improvements across a variety of NLP tasks.

Prompting is a technique where specific input phrases or instructions, known as "prompts", are used to guide the LLM's responses. Prompting leverages the model's pre-existing knowledge, allowing for immediate, on-the-fly task execution without the need for additional training. This method is highly flexible and efficient, especially for exploratory or dynamic tasks, as it does not require any modifications to the model's internal parameters. Previous work has demonstrated that techniques such as Chain-of-Thought (CoT) prompting~\cite{NEURIPS2022_9d560961} and few-shot learning~\cite{NEURIPS2020_1457c0d6} are highly effective in enhancing the performance of prompts. CoT prompting encourages the model to break down complex problems into intermediate reasoning steps, enabling more structured and logical responses. Few-shot learning involves providing the model with a small number of task-relevant examples within the prompt, which helps the model understand the specific format, context, or structure required for the task, even when it has not been explicitly fine-tuned for it. Drawing from various prompting techniques~\cite{10.1145/3560815, NEURIPS2022_9d560961, NEURIPS2023_271db992, yang2024large}, we designed a structured prompt framework to help us enhance computational efficiency in deploying LLMs for inference.

\section{Methodology}
\subsection{Overview}
\begin{figure}[!ht]
  \centering
  \includegraphics[width=\linewidth]{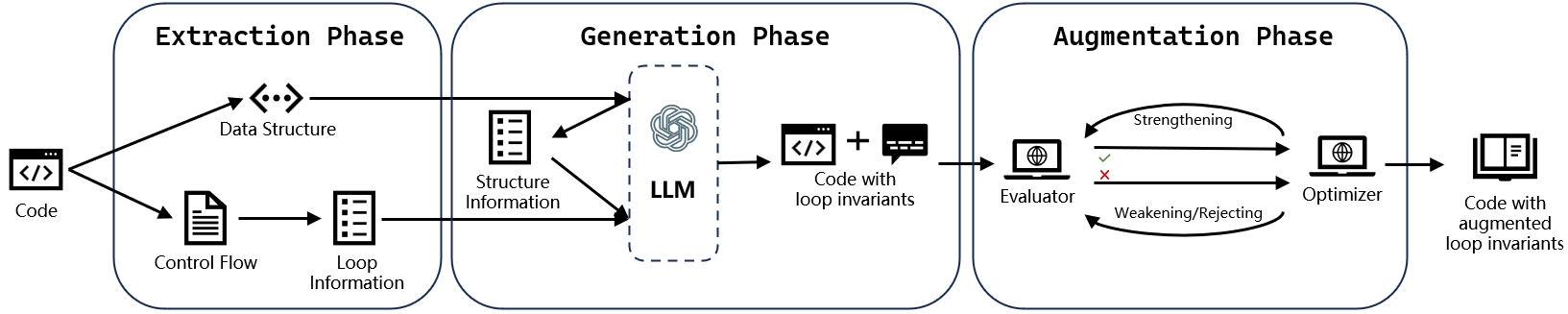}
  \caption{The Workflow of ACInv}
  \label{fig3:workflow}
\end{figure}

In Figure~\ref{fig3:workflow}, we present an overview diagram illustrating our proposed approach. To address the aforementioned challenge, we introduce ACInv, an LLM-based tool for automatic loop invariant generation. ACInv's process consists of three main phases: (1) \textbf{Extraction Phase}: ACInv initially extracts data structure and loop information from the target program. This involves performing basic control flow and value analyses, classifying loops, and gathering relevant variable information for each loop. (2) \textbf{Generation Phase}: During this phase, ACInv embeds the extracted data structure information into data structure prompts, querying LLMs to derive predicates that abstract key properties of the data structure. These predicates, along with loop-specific information, are then embedded into invariant prompts to generate loop invariants. If the current loop is not the first in the program, information from preceding loops, along with their generated invariants, will also be incorporated into the prompts. (3) \textbf{Augmentation Phase}: This section mainly consists of two components: the evaluator, and the optimizer, which are all composed of LLMs. The evaluator assesses the correctness of each generated invariant. For correct invariants, the optimizer attempts to strengthen them. For incorrect invariants, the optimizer weakens them to align with the correct program state. If an invariant cannot be refined to the correct level after $k$ iterations, it is discarded. In the remainder of this section, we will provide a detailed explanation of the entire methodology and its related intricacies.

\subsection{Problem description}

To provide a more formal description of this problem, we formulate the invariants generation task as follows: let $\mathbb{M}$ denote the LLMs, and $\mathbb{P}$ represent the program with a maximum of $n$ lines. Each line location in the code is denoted by $l_0, l_1, \cdots, l_n$, which may correspond to a line in a loop or data structure.

Let $d_{num}\in\mathcal{D}$ denote the indexed data structure instance retrieved by ordinal $num$, while $loop_{num}\in\mathcal{L}$ corresponds to an independent loop structure identified with the same indexing scheme, where $\mathcal{D}$ and $\mathcal{L}$ is the set of corresponding information. For each loop $loop_{num}$, assuming its specific position in the program $\mathbb{P}$ starts at line $l_i$, where $i\in\{1,n\}$, we define:
\begin{itemize}
    \item $\mathbf{P}$ as the prompts, which mainly include four prompts, predicate prompt as $\mathbf{P}_{ds}$, invariant prompt as $\mathbf{P}_{inv}$, evaluator prompt as $\mathbf{P}_{ev}$, and optimizer prompt as $\mathbf{P}_{op}$,
    \item $\mathbb{M}(\mathbf{P}(loop_{num}, l_i, \mathbb{P}, \mathcal{D}))$ as the query result specific to $(loop_{num}, l_i)$, and
    \item $\{I_k^{num}\}_{k=0}^{m_{num}}$ as the set of loop invariants at $l_i$.
\end{itemize}
Therefore, the current problem becomes: Given a program $\mathbb{P}$ with a maximum number of lines $n$, for each loop $loop_{num}\in\mathcal{L}$ and its corresponding line $l_i$ where $i\in \{1,n\}$, generate loop invariants $\{I_k^{num}\}_{k=0}^{m_{num}}\in\mathcal{I}$.

\begin{figure}[!ht]
  \centering
  \includegraphics[width=0.8\linewidth]{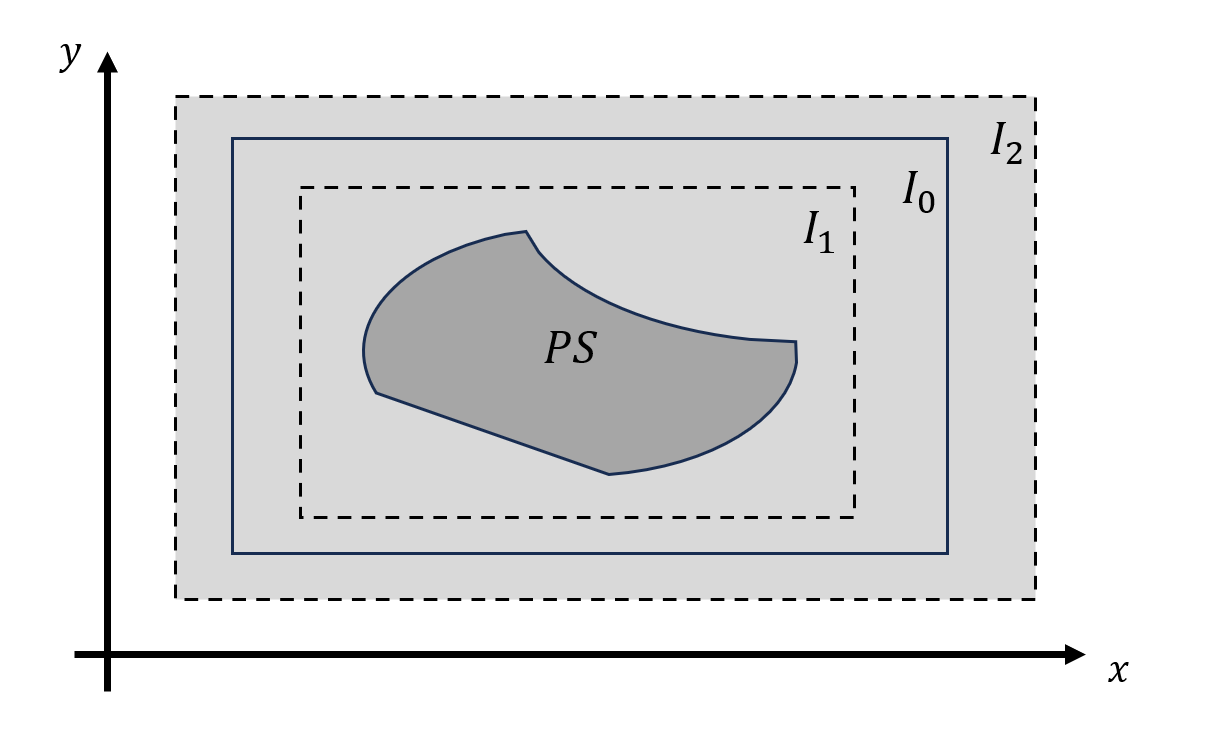}
  \caption{Diagram of properties}
  \label{fig:problem_properties}
\end{figure}

To formalize our requirements for the generated invariants' properties, we employ Fig.~\ref{fig:problem_properties} to illustrate key characteristics. In this representation, distinct variables correspond to different dimensions. The value ranges of these variables and their interrelationships collectively delineate all reachable program states, denoted as \(PS\) in the figure. For simplicity, only two variables \(x\) and \(y\) are shown; practical scenarios may involve higher-dimensional relationships.

The invariant \(I\) we aim to generate typically constitutes an \textit{over-approximation} of \(PS\). Consequently, our primary requirement is \textit{correctness}: the invariant must subsume all reachable program states, i.e., \(PS \subseteq I\). Under this correctness guarantee, a \textit{stronger} invariant is preferred, characterized by a tighter approximation closer to the true program state. Formally, smaller variable ranges yield stronger invariants. As illustrated, \(I_1\) is stronger than \(I_0\), whereas \(I_2\) is weaker.

\subsection{An illustrative example}
\begin{figure}[!ht]
  \centering
  \includegraphics[width=\linewidth]{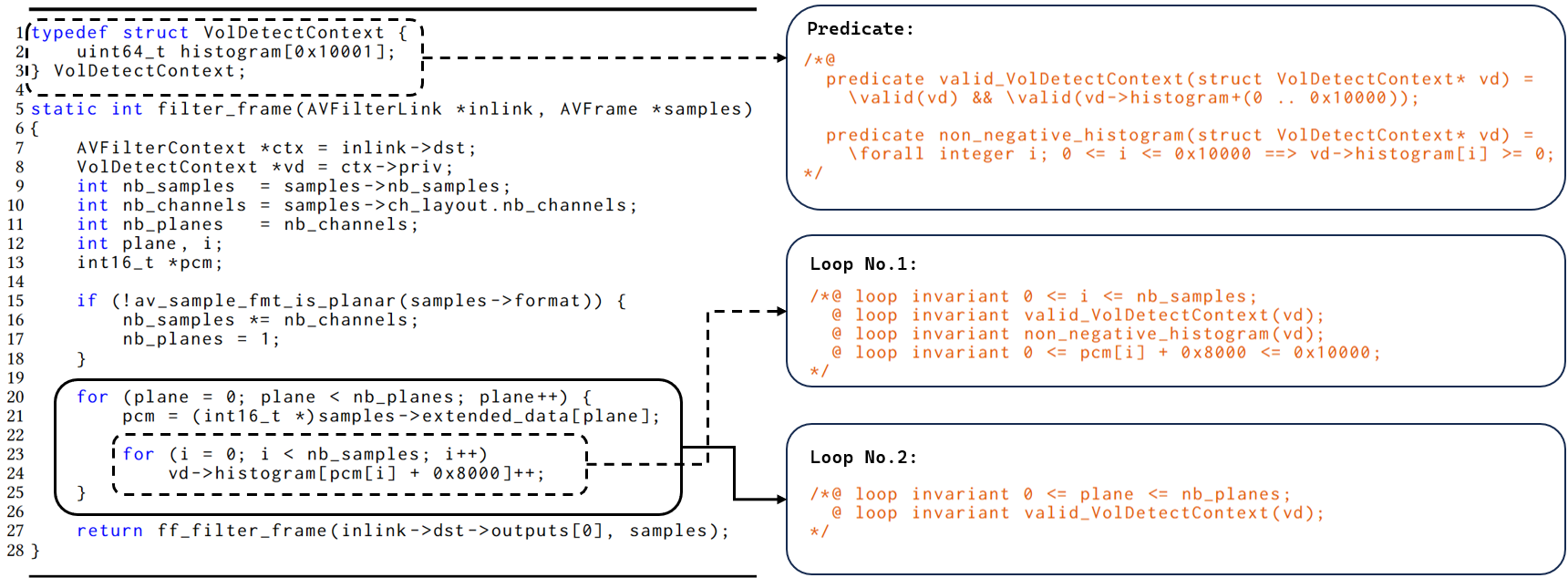}
  \caption{An illustrative example of C code from FFmpeg}
  \label{fig:illustrative example}
\end{figure}
Consider the example illustrated in Fig~\ref{fig:illustrative example}. This program fragment is from the widely-used FFmpeg~\cite{FFmpeg-repo} library, an open-source multimedia framework written in C that facilitates the recording, conversion, and streaming of audio, video, and other multimedia files. For this program fragment, we will focus on its main components, omitting less relevant details.

The \verb|VolDetectContext| structure is employed to store audio sample values for plotting frequency distributions. It contains a single member, \verb|histogram[]|, which is an array of unsigned 64-bit integers used to represent the distribution of audio sample frequencies. The \verb|filter_frame| function processes audio filters within FFmpeg, specifically handling audio frames and updating frequency distributions. It iterates through each audio plane of the input audio frame (\verb|AVFrame|) and adjusts the corresponding count in \verb|vd->histogram|, thus forming a nested loop structure. The expression \verb|pcm[i] + 0x8000| shifts the sample values (typically ranging from -32768 to 32767) to a 0 to 0xFFFF range for indexing purposes. This illustrative case forms the basis for our theoretical extension in the following section, where we derive generalized insights through rigorous analytical decomposition.

\subsection{Extraction Phase}
The primary objective of this phase is to extract structural and loop information from the target program $\mathbb{P}$. The input of this phase is a program or code segment $\mathbb{P}$, and output the data structure information $\mathcal{D}=\{d_{num}\}$ and loop information $\mathcal{L}=\{loop_{num}\}$. To achieve this, we implement a specialized parsing module that systematically traverses the entire program. The module first extracts data structure declarations and constructs a comprehensive data structure information list, denoted as $\mathcal{D}=\{d_{num}\}$. For each identified data structure, the module retrieves all its members and determines whether the data structure is recursive. A data structure is considered recursive if it contains a member of its own type. The recursiveness of a data structure significantly influences subsequent processing stages. At this phase, we merely flag the recursive nature of data structures, with detailed processing strategies to be elaborated in later stages.

More specifically, this stage actually achieves the following operations, such as Algorithm~\ref{static_analyse}:

\begin{algorithm}
\caption{The algorithm for extraction phase}
\label{static_analyse}
\SetKwInOut{Input}{Input}
\SetKwInOut{Output}{Output}
\Input{A C program $\mathbb{P}$}
\Output{Set $\mathcal{D}$ of data structures, Set $\mathcal{L}$ of loop information}
$\mathcal{D} \gets \emptyset$ \tcp*{Initialize data structure set} 
$\mathcal{L} \gets \emptyset$ \tcp*{Initialize loop information set}
$loopStack \gets \text{empty stack}$ \tcp*{Track nested loop positions}
\ForEach{node $n$ in the Abstract Syntax Tree (AST) of $P$}{
    \uIf{$n$ is a struct declaration}{
        $\mathcal{D} \gets \mathcal{D} \cup \{(datatype, dataname))\}$ \tcp*{Record struct}
    }
    \ElseIf{$n$ is a loop (for/while/do-while)}{
        push $n$ onto $loopStack$ \tcp*{Update loop position}
        $modifiedVars \gets \emptyset$ \;
        \ForEach{variable $v$ modified in $n$'s body}{
            $modifiedVars \gets modifiedVars \cup \{v\}$
        }
        $loopDepth \gets |loopStack|$
        $\mathcal{L} \gets \mathcal{L} \cup \{(modifiedVars, loopDepth)\}$ \tcp*{Record loop}
        pop $n$ from $loopStack$
    }
}
\Return ($\mathcal{D}$, $\mathcal{L}$)\;
\end{algorithm}

This algorithm performs static analysis of C programs through abstract syntax tree (AST) traversal, extracting structural and behavioral metadata. The procedure takes a C program $\mathbb{P}$ as input and outputs two sets: \(\mathcal{D}\) for data structure declarations and \(\mathcal{L}\) for loop information. Initialization establishes empty sets \(\mathcal{D}\) and \(\mathcal{L}\), along with an empty loop stack to track nesting depth. During AST traversal, when encountering a struct declaration, the algorithm extracts its composite data type definition \(\tau\) and adds \(\tau\) to \(\mathcal{D}\). For loop constructs (for/while/do-while), it pushes the loop node onto the stack, identifies all modified variables \(V_{\text{mod}}\) within the loop body, computes the current nesting depth as the stack size, records the tuple \((V_{\text{mod}}, \text{depth})\) to \(\mathcal{L}\), and finally pops the loop node from the stack. The algorithm terminates by returning the tuple \((\mathcal{D}, \mathcal{L})\), providing comprehensive structural definitions and loop mutation patterns essential for downstream program analysis tasks such as parallelization or dataflow verification, with \(\mathcal{O}(n)\) complexity proportional to AST node count.

Subsequently, ACInv systematically traverses the control flow graph of the program, with particular emphasis on loop structures. To facilitate structural analysis and provide clear decomposition pathways for the LLM, a hierarchical decomposition strategy is employed to break down complex loop constructs. During this process, ACInv identifies all loop instances within the program and assigns unique sequence numbers organized in a loop stack method. The numbering scheme follows these principles: for nested loops, the innermost loop receives the smallest number, while for parallel loops, priority is given to the loop appearing earlier in the program text. As illustrated in Fig~\ref{fig:illustrative example}, the innermost loop is labeled as 1, the enclosing loop as 2, and if there is a subsequent loop following this nested structure, it would be assigned number 3. This sequential numbering scheme establishes an analytical framework that guides the program analysis in subsequent phases, ensuring systematic exploration of the program's control flow.

Variable information can help LLM focus its attention on key parts, providing assistance in generating invariants. For each loop, ACInv also captures key variable information, which is specifically, all assigned variables and parameters passed as pointers in function calls. The parser will provide all variables that have undergone value changes and determine their variable types, including whether they belong to the abstract data types defined by the previously extracted data structure. All the information, along with the loop sequence number, forms the loop structure information $loop_{num}$ and is then maintained in the list $\mathcal{L}=\{loop_{num}\}$.

For the example in Fig~\ref{fig:illustrative example}, ACInv will record the declaration of the structure \verb|VolDetectContext|, provide a judgment that it is not recursive, and then traverse the program body. After completing the traversal of the loop body, it assigns sequence number 1 to the inner loop with line numbers 23-24 and extracts key variables: \verb|{i, vd}|; then assigns sequence number 2 to the outer loop with line numbers 20-25 and extracts key variables: \verb|{plane, pcm[]}|, as well as \verb|{i, vd}| inherited from the inner loop.

\subsection{Generation Phase}
In this phase, ACInv will leverage the information extracted in the previous phase to generate loop invariants by querying LLMs. The generation module of ACInv accepts the input $(\mathbb{P}, \mathcal{D}, \mathcal{L})$, which is the target program and the information of data structure and loop parsed in the extraction module, and give the output $\{{I^{\prime}}_k^{num}\}_{k=0}^{m}$ which is the preliminary invariant for each loop $loop_{num}\in\mathcal{L}$.

\begin{figure}[!ht]
  \centering
  \includegraphics[width=\linewidth]{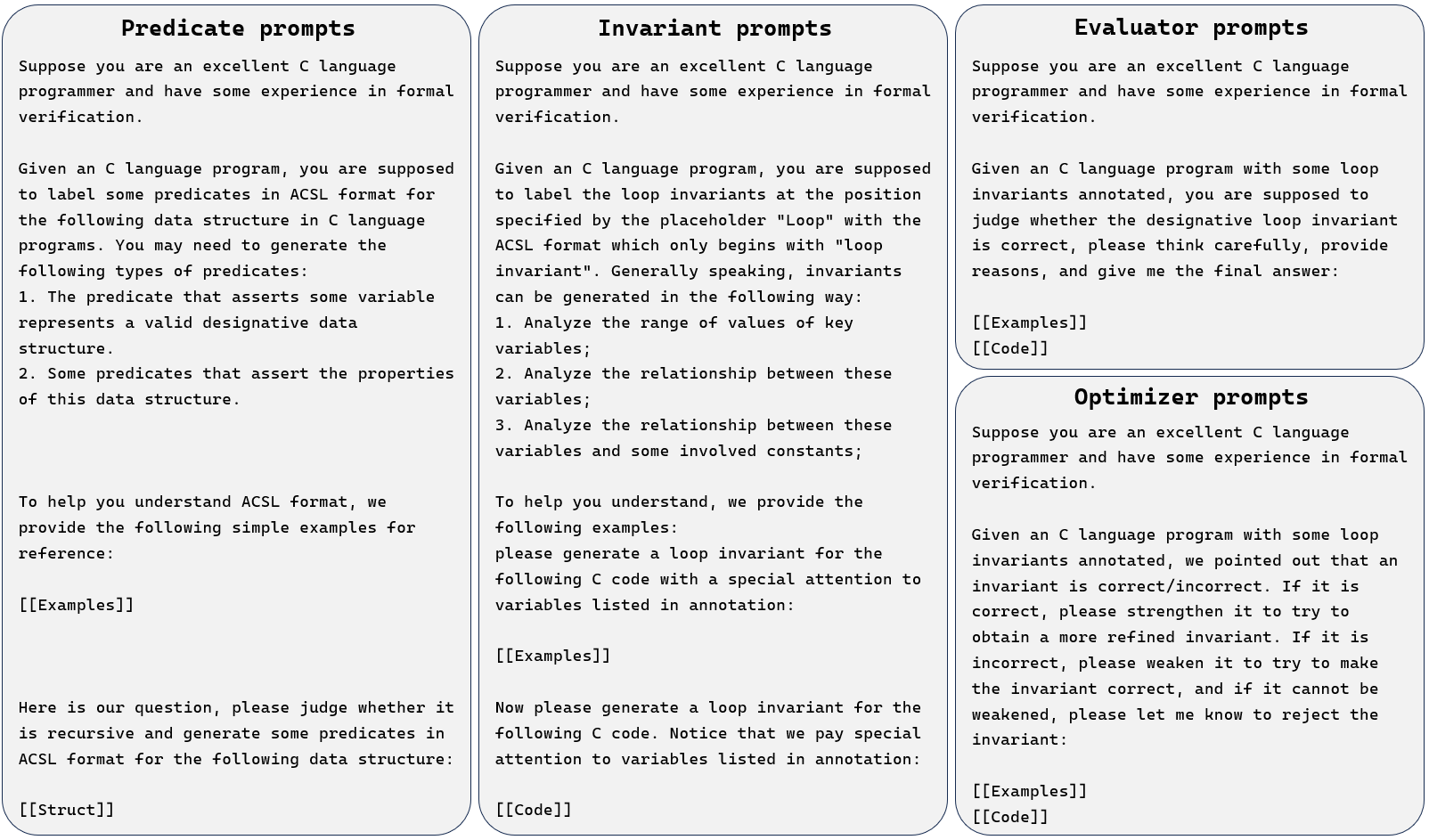}
  \caption{The prompts used in ACInv}
  \label{fig:prompts}
\end{figure}

In real-world programs, a fundamental challenge in automated invariant generation stems from the prevalence of custom data structures that frequently appear in real-world programs. These structures often represent either specialized variations of conventional data structures or, more commonly, intricate compositions tailored to specific application requirements. To address this challenge, our approach emphasizes the systematic characterization and formal specification of the properties of these data structures as a prerequisite step prior to invariant generation. This preliminary analysis enables more accurate and meaningful invariant derivation that properly reflects the semantic complexity of real-world data organizations. First, ACInv will embed the data structure information $\mathcal{D}= \{d_{num}\}$ into the predicate prompts $\mathbf{P}_{ds}$, which is illustrated in Figure~\ref{fig:prompts} and query the LLM to obtain predicate templates that may be used for this data structure. In prompts, we emphasize that these predicates need to be able to abstract the data structure itself, and in addition, provide some other properties about this data structure.

A critical observation reveals that the recursiveness of custom data structures constitutes a fundamental property that significantly influences both the formal specification of these structures in ACSL and the derivation of their associated properties. This characteristic plays a pivotal role in determining the appropriate representation and analysis of data structures throughout the verification process. In the syntax of ACSL, we can use \verb|predicate| to define predicates related to non-recursive data structures, and \verb|inductive| to define predicates related to recursive data structures. In order to help LLM understand predicate generation, we provided some examples to LLM, which are called shots in LLM prompting. The \verb|[[Examples]]| section in the predicate prompts acts as a placeholder that will be dynamically filled with specific examples. Similarly, the \verb|[[Code]]| placeholder will be replaced with the actual corresponding data structure. For non-recursively defined data structures, we provide a classic example: the stack, whose C language definition is usually as Figure~\ref{fig:stack}.

In order to help LLM understand predicate generation, we provided some examples to LLM, which are called shots in LLM prompting. For non-recursively defined data structures, we provide a classic example: the stack, whose C language definition is usually as Figure~\ref{fig:stack}.

\begin{figure}[!htbp]
    \centering
    \begin{subfigure}[c]{0.45\textwidth}
        \centering
        \begin{lstlisting}[style=cstyle]
struct stack_int{
    size_t top;
    int data[MAX_SIZE];
}
        \end{lstlisting}
        \caption{An non-recursive data structure example: stack}\label{fig:stack}
    \end{subfigure}
    \hfill
    \begin{subfigure}[c]{0.45\textwidth}
        \centering
        \begin{lstlisting}[style=cstyle]
struct Node{
    int data;
    struct Node * next;
};
        \end{lstlisting}
        \caption{A recursive data structure example: linked list}\label{fig:linked list}
    \end{subfigure}
    \caption{Two examples of data structure.}\label{fig:data structure}
\end{figure}

\begin{figure}[!htbp]
    \centering
    \subfloat[The predicates of stack\label{fig:stack_pred}]{
      \includegraphics[width=0.48\textwidth]{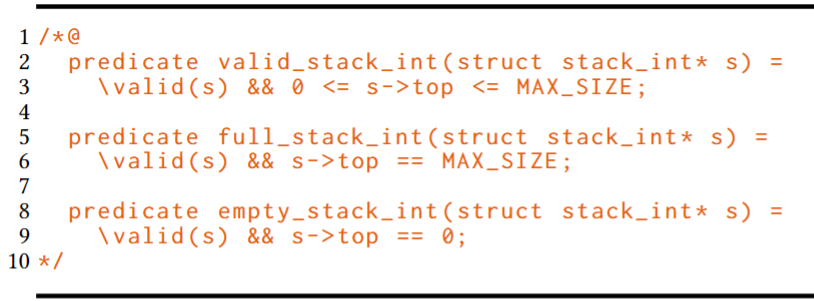}
    }
    \hfill
    \subfloat[The predicates of linked list\label{fig:linked_list_pred}]{
      \includegraphics[width=0.48\textwidth]{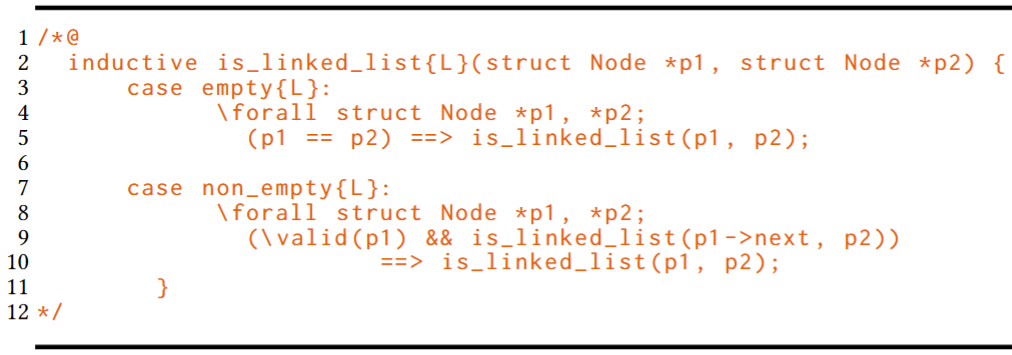}
    }
    \caption{Two examples of data structure.}\label{fig:ds_predicate}
\end{figure}

We expect the generated predicate to at least express that this is an effective stack; In addition, we also hope that it can express some other properties about the stack, such as whether the stack is empty or full. Therefore, in this example, we provide the following predicate such as Figure~\ref{fig:stack_pred}. In ACSL grammar, \verb|\valid(s)| means the pointer \verb|s| is valid to read or write. The recursive data structure is usually related to pointers. Taking inspiration from separation logic~\cite{1029817}, we provide such a recursive definition for the following single linked list, which considered both empty and non-empty linked lists, and expressed that the pointer \verb|p1| to \verb|p2| is a single linked list such as Figure~\ref{fig:linked_list_pred}.

Then, combined with the possible predicates, ACInv embeds the program $\mathbb{P}$ and the loop information $\mathcal{L}$ into the invariants prompt $\mathbf{P}_{inv}$ to query the LLM. Based on the order of the loops, ACInv will sequentially query the LLM for the loop with the smallest ordinal number, and require the LLM to generate invariants for that loop. Within each loop, we embed the processed program into prompts and require the LLM to analyze the following: (1) the range of values of the annotated variables; (2) Possible logical relationships between these variables and constants; (3) Does the data structure variable satisfies the predicate. The \verb|[[Example]]| section provides multiple program instances accompanied by their corresponding loop invariants, with particular emphasis on detailed explanations of nested loop structures and the associated data structures utilized within these examples. This section serves to illustrate both the general patterns and specific considerations for invariant generation in complex control flow scenarios and custom data structure implementations. From the responses generated by the LLM, we systematically extract the invariants corresponding to their designated positional indices. Specifically, for a given loop identified by its numerical label $num$, we obtain its corresponding invariant $\{{I^{\prime}}_k^{num}\}_{k=0}^{m}$. When the traversal of all loops is completed, we obtain a comprehensive set of invariants, denoted as $\mathcal{I}^{\prime}=\{\{{I^{\prime}}_k^{1}\}_{k=0}^{m_1}, \{{I^{\prime}}_k^{2}\}_{k=0}^{m_2}, \cdots, \{{I^{\prime}}_k^{num}\}_{k=0}^{m_{num}}, \cdots\}$.

Figure~\ref{fig:illustrative example} illustrates the majority of the process for this step. Initially, two predicates are generated based on the data structure, which represents whether it is a valid \verb|VolDetectContext| and the non-negativity of the array \verb|histogram[]|. Using these predicates, ACInv first derives an invariant for Loop 1, which is an inner loop. Once this invariant is obtained, all invariants within this loop are forwarded to the next stage for enhancement. Following this, the enhanced invariant is used as a reference for generating the invariant for Loop 2 by querying the LLM.

\subsection{Augmentation Phase}
A prevalent challenge is that the invariants generated by LLMs often tend to be weak—for instance, trivial solutions such as $\mathcal{I} = TRUE$ are frequently produced. To address this issue while simultaneously filtering out incorrect candidates, we introduce an augmentation phase. In this phase, ACInv focuses on refining the generated invariants by either strengthening or weakening them. Building upon the previous phase where we derived the initial set of invariants $\mathcal{I}^{\prime}$ from program $\mathbb{P}$, we now aim to perform a deeper analysis to refine and enhance the invariant collection. We seek to obtain an optimized set of invariants $\mathcal{I}$ that more accurately captures the program's behavioral properties and provides stronger verification guarantees. Unlike the previous phase, this stage is loop level, which means that for each individual loop, the generated invariant will first be sent to this stage to try to be enhanced, and then the enhanced invariant group will be sent back to the generation phase as a reference for generating the invariant in subsequent loops. For each loop $loop_{num}$, the $i$-th generated invariant ${I^{\prime}}^{num}_i$ will be enhanced independently.

It begins by utilizing an evaluator to assess the correctness of the invariants and then calls upon an optimizer to take appropriate actions based on the evaluation. Both the evaluator and optimizer are LLM-based. As illustrated in Fig~\ref{fig:prompts}, we use elaborately designed dedicated prompts and examples to implement the LLM-based evaluator and optimizer. It is important to note that these components utilize LLMs independently, both with each other and with the LLM employed in the aforementioned generation phase. As mentioned earlier, we hope to use evaluators to ensure the correctness and optimizers to guarantee a stronger invariant. However, a very normal concern is that the LLM may make mistakes. However, relevant studies~\cite{chakraborty-etal-2023-ranking} have shown that the existing LLMs already have a strong enough reasoning ability and can give results with relatively reliable correctness rates to a certain extent. Through multiple independent information processing operations, we can further reduce its error probability. We expect to complete an end-to-end invariant generation work based on LLM, and thus it can be applied to scenarios that tolerate certain errors.

If an invariant is deemed correct, ACInv will attempt to "strengthen" it. In the context of invariants, this typically involves narrowing the range of program variable states represented, resulting in a more precise invariant. Conversely, if the invariant is deemed to be incorrect, ACInv seeks to "weaken" it. This process involves expanding the range of program variable states to encompass the correct set of states within the revised invariant. By broadening the scope of the incorrect invariant, ACInv aims to adjust it to better reflect the valid program states. However, our weakening process incorporates a rejection mechanism, allowing the LLM to decline the weakening attempt if it determines that the invariant cannot be properly weakened while maintaining correctness. Following this step, the results are subsequently returned to the evaluator for further validation. To control the iterative refinement process, we introduce a parameter $k$ that limits the maximum number of query cycles. When this threshold is reached, invariants that remain incorrect are permanently rejected, while those that have been successfully validated are retained in the final invariant set. At this stage, we obtain the final set of invariants, denoted as $\mathcal{I}=\{\{I_k^{1}\}_{k=0}^{m_1}, \{I_k^{2}\}_{k=0}^{m_2}, \cdots, \{I_k^{num}\}_{k=0}^{m_{num}}, \cdots\}$.

In practice, before entering the augmentation phase, the invariant generated by ACInv for Loop 2 in Figure~\ref{fig:illustrative example} is \verb|0 <= plane < nb_planes|. However, during the execution of the loop, although the outer loop condition is \verb|plane <  nb_planes|, the final iteration may satisfy \verb|plane == nb_planes|. Thus, this invariant does not cover all possible program states. The evaluator can identify this issue and the optimizer can adjust the invariant by expanding its range to \verb|0 <= plane <= nb_planes|, ultimately providing the refined invariant.

\section{Experiment}
\subsection{Reaserch Question}
In this section, we will propose several intriguing research questions concerning ACInv and conduct experiments to address these questions:
\begin{itemize}
    \item \textbf{RQ1}: Can ACInv generate correct predicate templates for complex programs with data structure and thus generate correct loop invariants?
    \item \textbf{RQ2}: Can ACInv generate correct loop invariants for complex programs without data structure?
    \item \textbf{RQ3}: How useful is the augmentation phase for ACInv?
\end{itemize}

\subsection{Experimental Setup}
\textbf{Benchmark} Our dataset is mainly divided into two parts: those with data structures and those without data structures. The dataset of data structure consists of three parts: LIG-MM~\cite{liu2024generalloopinvariantgeneration, LIG-MM-repo}, SV-COMP~\cite{10.1007/978-3-031-57256-2_15, SV-comp-24-repo}, and FFmpeg~\cite{FFmpeg-repo}. We will elaborate on these three datasets in more detail later. The dataset without data structure mainly comes from the benchmark from AutoSpec~\cite{10.1007/978-3-031-65630-9_16}, and we will also briefly describe this dataset.

First, we selected the benchmark from LIG-MM, which is a new benchmark specifically for programs with complex data structures and memory manipulations. LIG-MM is also divided into four parts: (1) the course programs which originate from course homework; (2) SV-COMP~\cite{10.1007/978-3-030-99527-0_20}, the competition on software verification; (3) Benchmark from SLING~\cite{10.1145/3314221.3314634}, which is a separation logic loop invariant inferring tool using dynamic analysis; (4) Some real-world programs from well-known software and systems. LIG-MM consists of C code and focuses on some complex data structures, especially the single-linked list, double-linked list, tree, and hash table, which can be an appropriate reference dataset. Due to the limit on the number of tokens, we selected 227 examples with relatively few lines of code for testing.

We also picked C programs from SV-COMP 2024. LIG-MM and the benchmark of AutoSpec both select some programs from SV-COMP. But unlike them, we chose different parts. We selected twenty-six programs, all of which include at least one loop and one data structure. Among these data structures, we excluded all structures containing the single-linked list, double-linked list, tree, and hash table to avoid duplication with the selected parts in LIG-MM. Therefore, these programs also include many different types of structures.

To evaluate the effectiveness of ACInv on real-world programs, we selected several program segments from the widely used FFmpeg, the audio and video processing library. Given the complexity of function calls and structures within C programs, as well as the extensive codebase, we focused on specific portions of the program. We applied simple partitioning techniques, ensuring that ACInv could concentrate on the relevant functions for analysis, and we picked the programs with at least one loop and at least one data structure.

We also selected programs from AutoSpec, an automated tool for generating ACSL specifications for C programs. AutoSpec's benchmark assesses the tool’s ability to synthesize specifications, including loop invariants, preconditions, and postconditions of functions. Notably, some C programs in this benchmark do not contain loops, so we specifically chose a subset that includes loops for our testing.

Benchmark from AutoSpec comprises four parts, which we introduce briefly: (1) Frama-C problems\cite{frama-c-problem-repo}, a repository focused on program verification challenges for Frama-C. We selected a portion containing C programs with loops. (2) SyGuS\cite{alur2019syguscomp2018resultsanalysis}, which represents the invariants track from the Syntax-Guided Synthesis competition, featuring only programs with linear loop structures. (3) OOPSLA-13\cite{10.1145/2544173.2509511, 10.1145/2509136.2509511}, and (4) SV-COMP\cite{10.1007/978-3-030-99527-0_20}, both of which include benchmarks that may have nested or multiple loops. All these benchmarks consist of C code, primarily focused on numerical programs without explicit structural declarations.

\begin{table*}[!ht]
    \centering
    \caption{Comparison of Performance on Multiple Datasets}
    \fontsize{7pt}{9.5pt}\selectfont
    \renewcommand{\arraystretch}{1.3}
    \newcommand{\colwidth}{1cm}
    \setlength{\tabcolsep}{3pt}
    \begin{tabularx}{\textwidth}{p{1.5cm}>{\centering\arraybackslash}p{\colwidth}>{\centering\arraybackslash}p{\colwidth}>{\centering\arraybackslash}p{\colwidth}>{\centering\arraybackslash}p{\colwidth}>{\centering\arraybackslash}p{\colwidth}>{\centering\arraybackslash}p{\colwidth}>{\centering\arraybackslash}p{\colwidth}>{\centering\arraybackslash}p{\colwidth}>{\centering\arraybackslash}p{\colwidth}>{\centering\arraybackslash}p{\colwidth}}
        \toprule
        \multirow{2}{*}{\textbf{Tools}} & \multicolumn{3}{c}{\textbf{Data Structure}} & \multicolumn{4}{c}{\textbf{non-Data Structure}} & \multicolumn{3}{c}{\textbf{Overall}} \\
        \cmidrule(lr){2-4} \cmidrule(lr){5-8} \cmidrule(lr){9-11}
        & LIG-MM & SVCOMP24 & FFmpeg & Frama-C & SyGuS & OOPSLA & SVCOMP22 & Struc & non-Struc & All \\
        \midrule
        \textbf{non-LLM} & & & & & & & & & & \\
        SLING & .171 & .000 & .000 & - & - & - & - & .128 & - & .055 \\
        CODE2Inv & - & - & - & .000 & .549 & .196 & .000 & - & .358 & .204 \\
        CLN2Inv & - & - & - & .000 & \cellcolor{gray!30}\textbf{.932} & .000 & .000 & - & .541 & .309 \\
        \midrule
        \textbf{LLM-based} & & & & & & & & & & \\
        LLM-SE & \cellcolor{gray!30}\textbf{.419} & - & - & - & - & - & - & .314 & - & .135 \\
        GPT-4o & .054 & .192 & .059 & .172 & .278 & .261 & .143 & .076 & .249 & .175 \\
        AutoSpec & .016 & .308 & .059 & .414 & .857 & \cellcolor{gray!30}\textbf{.826} & \cellcolor{gray!30}\textbf{.762} & .064 & \cellcolor{gray!30}\textbf{.786} & .476 \\
        ACInv-3.5 & .139 & .307 & .118 & .345 & .714 & .457 & .429 & .163 & .589 & .407 \\
        ACInv-4o & .318 & \cellcolor{gray!30}\textbf{.462} & \cellcolor{gray!30}\textbf{.353} & \cellcolor{gray!30}\textbf{.483} & .797 & .761 & \cellcolor{gray!30}\textbf{.762} & \cellcolor{gray!30}\textbf{.343} & .747 & \cellcolor{gray!30}\textbf{.574} \\
        \bottomrule
    \end{tabularx}
    \label{tab:Comparison of Performance}
\end{table*}

\textbf{Baseline} In this study, we compare the performance of several tools with ACInv on datasets both with and without data structures. For datasets with data structures, we evaluate SLING, LLM-SE~\cite{liu2024generalloopinvariantgeneration}, GPT-4o, and AutoSpec. SLING is a dynamic analysis tool for inferring loop invariants based on separation logic. LLM-SE leverages a fine-tuned LLM (which is CodeGen~\cite{nijkamp2023codegen2lessonstrainingllms} in the specific implementation) to generate separation logic loop invariants, incorporating predicate decomposition into its training. AutoSpec employs hierarchical specification generation to produce ACSL-style annotations. Both LLM-SE and AutoSpec utilize verification checkers to ensure the correctness of the generated invariants. GPT-4o, in this context, is used by directly embedding the code into prompts, asking GPT-4o-2024-08-06 to generate loop invariants without employing specific prompting techniques or examples.

For datasets without data structures, we assess CODE2Inv~\cite{NEURIPS2018_65b1e92c, 10.1007/978-3-030-53291-8_9}, CLN2Inv~\cite{Ryan2020CLN2INV:}, GPT-4o, and AutoSpec. CODE2Inv and CLN2Inv are state-of-the-art invariant generation tools for numerical programs, based on graph neural networks and continuous logic networks, respectively. AutoSpec was originally a tool for testing on this benchmark, and we naturally referred to its results.

\textbf{Environment} We primarily employ \textit{GPT-4o-2024-08-06} as the core model for implementing ACInv, with \textit{GPT-3.5-turbo} serving as a control in our experiments. The experiments are performed on a system equipped with an AMD Ryzen 9 5950X CPU, running Ubuntu 23.04, 128GB of RAM, and two NVIDIA RTX 4090 GPUs, each with 24GB of memory.

\subsection{RQ1: Performance on Structure Datasets}
Table~\ref{tab:Comparison of Performance} shows the main results of our experiments, where completion means the rate of programs whose loop invariants can be inferred correctly to complete the verification of this program. In this experiment, considering the balance between time and performance, we set the augmentation iteration parameter k to 5. Compared to the traditional tool SLING, LLM-based technologies, including ACInv, demonstrate strong applicability. While SLING can only reason about the linked list portion of LIG-MM, it struggles with other complex data structures. Experimental results from LLM-SE indicate good performance on singly linked lists (sll), doubly linked lists (dll), trees, and hash tables. However, LLM-SE did not provide training sets or fine-tuned models, limiting the ability to evaluate its performance on other data structures. Additionally, LLM-SE incorporates pre-defined predicates during fine-tuning, enabling abstraction during reasoning, though this approach may restrict its effectiveness on other data structures or non-pointer data structures.

In comparison to pure LLM querying and the latest technology, AutoSpec, ACInv performs well on datasets involving data structures. This performance is anticipated. In pure LLM querying, we observed that while the LLM can recognize data structures in the program and analyze the associated variables, it struggles to abstract them into suitable logical forms. This often results in accurate formulas being expressed in natural language rather than logical terms. Regarding AutoSpec, its examples primarily focus on numerical types, leading to suboptimal performance on benchmarks involving data structures.

The FFmpeg dataset presents significant challenges due to its many unpredictable user-defined data structures, making it difficult for large models to abstract their properties. None of the tools performed well in this scenario, though ACInv exhibited some advantages. Overall, the experiments suggest that ACInv possesses a notable degree of abstraction and reasoning capability across various datasets involving complex data structures.

\subsection{RQ2: Performance on non-Structure Datasets}
In numerical programs, ACInv demonstrated performance comparable to AutoSpec. The SyGuS benchmarks, which consist of relatively simple programs with a single loop and focus on basic numerical transformations, saw CLN2Inv achieving the best results. However, both CODE2Inv and CLN2Inv performed poorly on other datasets, likely due to limitations in their training sets, highlighting the broader applicability of LLM-based technologies. The Frama-C problems benchmark, which involves nested loops and more complex variable interactions, posed greater challenges. ACInv, leveraging value analysis, performed notably better in these cases. 

Overall, ACInv approached state-of-the-art performance in numerical programs without data structures and demonstrated clear advantages across diverse datasets.

\subsection{Case Study:}

\begin{figure}
    \centering
    \includegraphics[width=0.5\linewidth]{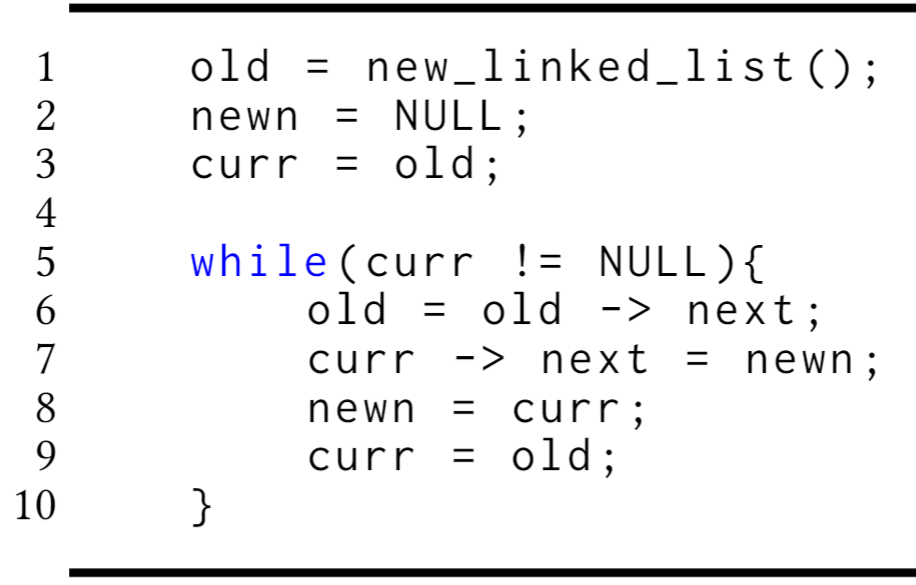}
    \caption{In-place reversal of the singly linked list}
    \label{fig:sll_rev}
\end{figure}

Considering the example as Figure~\ref{fig:sll_rev}. This program snippet performs an in-place iterative reversal of a singly linked list through pointer manipulation. Three pointers are initialized: \texttt{old} references the head of the original list, \texttt{newn} is set to \texttt{NULL}, representing the nascent reversed list, and \texttt{curr} is initialized to traverse the original list starting at \texttt{old}. During each iteration, \texttt{old} is advanced to its successor node to preserve traversal integrity, the \texttt{next} pointer of \texttt{curr} is redirected to \texttt{newn} to reverse linkage direction, \texttt{newn} is updated to reference the current node, thus extending the reversed sublist, and \texttt{curr} is reset to \texttt{old} for subsequent processing. The loop terminates when \texttt{curr} becomes \texttt{NULL}, at which point \texttt{newn} references the head of the fully reversed list. To give a specific example, if this linked list was originally \texttt{A->B->C->NULL}, it becomes \texttt{C->B->A->NULL} after the reversal.

Most of the other methods based on LLM can only obtain one invariant for this loop, namely \texttt{old == curr}. By observing the detailed output of the LLM, we can find that the LLM can actually capture some more specific properties. That is, this linked list is divided into two parts from the pointers old and curr. The part before curr is the flipped linked list, while the part after old is the linked list that has not been flipped yet. However, LLMs cannot describe this property completely formally. Instead, it describes it in a semi-formal and semi-natural language way, such as \textit{\texttt{curr == suffix} of the original list starting at position \texttt{i}.} And our work can provide a more formal description of this nature, namely \texttt{is\_linked\_list(curr,NULL)}, \texttt{is\_linked\_list(old,NULL)}.

\subsection{RQ3: Testing for the Augmentation}
\begin{figure}[!ht]
    \centering
    \includegraphics[width=0.48\textwidth]{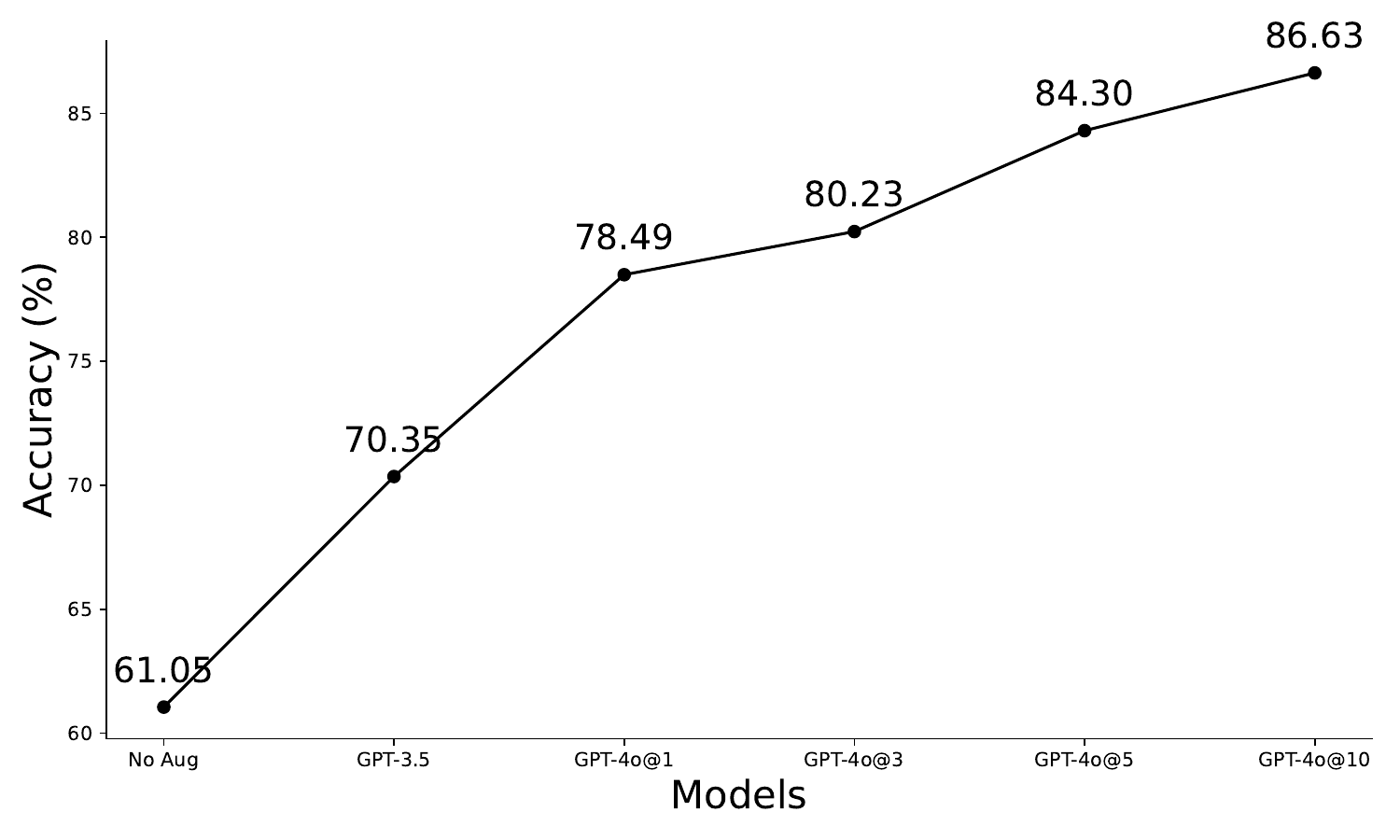}
    \caption{Performance of Augmentation of ACInv}
    \label{fig:performance_augmentation}
\end{figure}
Figure~\ref{fig:performance_augmentation} shows the results of the experiment on the augmentation of ACInv. We tested the invariants generated by ACInv on a structured dataset, where Acc means the proportion of generated invariants that can pass the frama-c check, and model@k means k augmentation iterations were performed on the model. The experiment shows that when we use appropriate prompts and examples, the augmentation stage can judge the correctness of the invariants with a relatively high probability and filter out incorrect invariants.

\subsection{Threats to validity}
\textbf{Data leakage}. The extensive training sets used in LLMs can lead to potential data leakage, meaning that some of the code related to the questions we inquired about might have appeared in the training data of certain models. Although we have applied simple variable renaming and reordered the code, similar issues may still arise.

\textbf{Correctness determination}. Most related work, particularly during the final evaluation stage (and in some cases even in earlier stages), employs a deterministic checker to validate the generated invariants for filtering purposes. We did not adopt this approach here, partly due to performance overhead concerns and partly to avoid making the generated invariants dependent on a specific checker. In our experiments, we observed that some correct invariants could not be verified by the checker due to limitations in the checker's capability and available computational resources. In future work, we aim to follow up on this aspect, seeking to strike a more effective balance between correctness and scalability.

\section{Related Work}
\subsection{LLM for static analysis}
Recent advances in LLMs have opened new avenues in static analysis. By integrating LLMs with traditional static analysis techniques, their scope can be broadened, leading to more effective and deeper program comprehension.

LLM4SA~\cite{10.1145/3653718} leverages LLMs to analyze static warnings, reducing the need for manual effort. LLift~\cite{10.1145/3649828} integrates static analysis with LLM-based prompting, successfully identifying Use Before Initialization (UBI) bugs in the Linux kernel. CFStra~\cite{10.1007/978-3-031-64626-3_22} employs LLMs to automatically select verification strategies based on code features and specifications. IRIS~\cite{li2024llmassistedstaticanalysisdetecting} combines LLMs with static analysis to perform whole-repository reasoning, detecting security vulnerabilities. NS-Slicer~\cite{10.1145/3649814} uses a fine-tuned model to predict static program slices for both complete and partial code. Codeplan~\cite{10.1145/3643757} synthesizes a multi-step chain of edits to assist with repository-level coding. ClarifyGPT~\cite{10.1145/3660810} enhances code generation by leveraging LLMs, while AutoFL~\cite{10.1145/3660771} utilizes LLMs to generate explanations of bugs for the purpose of Fault Localization (FL). Overall, incorporating LLMs has mitigated challenges related to applicability and performance overhead in static analysis and has enhanced the ability to analyze real-world code.

\subsection{Invariants Generation}

\textbf{Symbolic methods} Traditional symbolic methods for invariant generation typically involve direct code analysis, mainly including abstraction interpretation and constraint solving. Abstract interpretation~\cite{10.1145/512950.512973} approximates a program's behavior by mapping its concrete semantics to a simplified abstract domain, often using fixed-point calculations to abstract program states in the presence of loops. Enric et al.\cite{10.1007/978-3-540-27864-1_21} developed a framework for generating conjunctions of loop invariants constrained to linear inequalities or polynomial equations, later extending the approach to polynomial invariants of bounded degree\cite{RODRIGUEZCARBONELL200754}. Steven et al.~\cite{10.1007/978-3-319-46520-3_30} applied a combination of Groebner bases, linear algebra, and abstract interpretation to generate polynomial invariants. Constraint-based methods, on the other hand, rely on pre-defined templates and employ SMT solvers to filter out appropriate invariant expressions. Farkas' lemma~\cite{10.1007/978-3-540-27864-1_7, 10298567, 10.1007/978-3-031-64626-3_19, 10.1007/978-3-540-45069-6_39, 10.1145/3563295} is frequently used to extract linear relationships from loops, while other constraint templates, such as those based on Craig’s interpolant~\cite{10.1007/978-3-540-45069-6_1, 10.1145/2544173.2509511, 10.1145/2509136.2509511}, are also employed.

Separation Logic (SL)~\cite{1029817} is an extension of Hoare logic to reason about memory structures and the interactions of programs with dynamically allocated memory. It introduced the concept of "separating conjunction," which allows for concise reasoning about disjoint memory regions and parallel computation. Based on the extension of list manipulation~\cite{Magill2005InferringII}, THOR~\cite{10.1007/978-3-540-70545-1_41, 10.1145/1706299.1706326, 10.1145/1707801.1706326} and MemCAD~\cite{10.1007/978-3-319-57288-8_15} are able to infer loop invariants on numeric and linked list programs. More recently, SLING~\cite{10.1145/3314221.3314634} advances SL invariant generation by employing dynamic analysis to track memory and numerical changes in pointer-related structures.

\textbf{Learning-based methods} Learning-based methods often build upon template-based approaches by learning the parameters within these templates through various machine-learning techniques and ultimately invoking an SMT solver to filter suitable invariants. C2I~\cite{10.1007/978-3-319-08867-9_6} employs a randomized search to discover candidate invariants, followed by a checker to validate them. ICE~\cite{10.1007/978-3-319-08867-9_5} did learning for synthesizing invariants, leveraging examples, counterexamples, and implications. Then they combined them with decision trees~\cite{10.1145/2914770.2837664}. CODE2INV~\cite{NEURIPS2018_65b1e92c, 10.1007/978-3-030-53291-8_9} utilizes reinforcement learning and graph neural networks to learn program structures. LIPuS~\cite{10.1145/3597926.3598047} also adopts reinforcement learning and uses pruning to reduce the search space. CLN2INV~\cite{Ryan2020CLN2INV:} introduces Continuous Logic Networks (CLN) to automatically learn loop invariants directly from program execution traces, while Gated-CLN (G-CLN)~\cite{10.1145/3385412.3385986} extends this approach to allow the model to robustly learn general invariants across a large number of terms.

\textbf{LLMs for loop invariants} Just as we talked about above in the experiment section, LIG-SE~\cite{liu2024generalloopinvariantgeneration} fine-tuned the large model and combined it with a checker, giving the LLM the ability to handle separation logic. It can also handle data structures and abstract predicate properties to a certain extent. Autospec~\cite{10.1007/978-3-031-65630-9_16} employs hierarchical specification generation to produce ACSL-style annotations, which include loop invariants and have achieved good performance in the processing of numerical programs. Both LLM-SE and AutoSpec call the checker at the end to ensure the correctness of the generated invariants. Several other works have explored the intersection of LLMs and loop invariant generation with promising outcomes. The iRank tool~\cite{chakraborty-etal-2023-ranking} introduces a re-ranking approach that evaluates loop invariants produced by LLMs. Unlike other methods, iRank does not focus on generating loop invariants but instead addresses the challenge of verifying the correctness of LLM-generated invariants. To achieve this, the authors employed two distinct LLM-based embedding models to learn the criteria for determining invariant correctness. Kexin et al.\cite{pmlr-v202-pei23a} aimed to fine-tune a pre-trained LLM for abstract reasoning over program executions, allowing for the generation of loop invariants. Their approach relies on a dynamic analyzer to trace program executions, which are then used to derive invariants. However, their evaluation is limited to a comparison against invariants generated by the dynamic analyzer, which poses inherent constraints. Loopy\cite{kamath2023findinginductiveloopinvariants}, on the other hand, continuously queries LLMs to generate invariants and employs a verification tool to ensure correctness. While they enhance the generated invariants by cyclically applying the Houdini algorithm and making repairs, the repeated invocation of the verifier introduces significant performance overhead.

\section{Conclusion}
In this paper, we proposed ACInv, an Automated Complex program loop Invariant generation tool leveraging Large Language Models (LLMs) in conjunction with static analysis. ACInv can abstract the data structure properties in complex programs and combine them to generate invariants. ACInv can also evaluate and enhance the generated invariants to a certain extent. We have conducted experiments on the effect of ACInv, and the experiments show that ACInv has achieved certain results on various data sets.

\subsubsection*{Acknowledgment}
This work is supported by the National Natural Science Foundation
of China Grant No. 61872232.

\bibliographystyle{elsarticle-num-names} 
\bibliography{ref}

\end{document}